\documentclass[aps,twocolumn,nofootinbib,showpacs,superscriptaddress,hyperref]{revtex4-1}

\usepackage{amsfonts}
\usepackage{amsmath}
\usepackage{amssymb}
\usepackage{bm}
\usepackage{dcolumn}
\usepackage{graphicx}
\usepackage{graphics}
\usepackage[latin1]{inputenc}
\usepackage{latexsym}
\usepackage{rotating}
\usepackage{hyperref}
\usepackage[all]{hypcap} 
\usepackage{xspace} 
\usepackage[usenames]{color}
\usepackage{mathrsfs}

\usepackage{ulem}
\definecolor {darkgreen}{rgb}{0.2,0.7,0.2}

\newcommand\be{\begin{equation}}
\newcommand\ba{\begin{eqnarray}}
\newcommand\ee{\end{equation}}
\newcommand\ea{\end{eqnarray}}
\newcommand\bw{\begin{widetext}}
\newcommand\ew{\end{widetext}}

\begin{document}
\title{Improved Universality in the Neutron Star Three-Hair Relations}    

\author{Barun Majumder}
\affiliation{Department of Physics, Montana State University, Bozeman, MT 59717, USA}
\affiliation{Department of Physics, IIT Gandhinagar, Ahmedabad, India}

\author{Kent Yagi}
\affiliation{Department of Physics, Montana State University, Bozeman, MT 59717, USA}

\author{Nicol\'as Yunes}
\affiliation{Department of Physics, Montana State University, Bozeman, MT 59717, USA}

\begin{abstract} 

No-hair like relations between the multipole moments of the exterior gravitational field of neutron stars have recently been found to be approximately independent of the star's internal structure. 
This approximate, equation-of-state universality arises after one adimensionalizes the multipole moments appropriately, which then begs the question of whether there are better ways to adimensionalize the moments to obtain stronger universality. 
We here investigate this question in detail by considering slowly-rotating neutron stars both in the non-relativistic limit and in full General Relativity. 
We find that there exist normalizations that lead to stronger equation-of-state universality in the relations among the moment of inertia and the quadrupole, octopole and hexadecapole moments of neutron stars.  
We determine the optimal normalization that minimizes the equation-of-state dependence in these relations.
The results found here may have applications in the modeling of X-ray pulses and atomic line profiles from millisecond pulsars with NICER and LOFT.

\end{abstract}

\pacs{04.25.Nx,97.60.Jd}
\date{\today}
\maketitle

\section{Introduction}

One of the most interesting physical results that one may derive from neutron star (NS) observations is a better understanding of the supra-nuclear equation of state (EoS), i.e.~the relation between internal density and pressure at densities beyond nuclear~\cite{lattimer_prakash2001,lattimer-prakash-review,Lattimer:2012nd}. Imagine, for example, that one were to observe the NS mass and its radius \emph{independently}. Since the relation between the mass and radius of a NS is highly sensitive to the EoS, such observations would place strong constraints on the latter. But the NS radius, in particular, is currently very hard to measure with sufficient accuracy, which makes constraints on the EoSs not sufficiently strong~\cite{steiner-lattimer-brown,Lattimer:2013hma}.

Ignorance of the EoS can have a strong impact on the amount of information that can be extracted from astrophysical observations. When the EoS is unknown, more model parameters are typically needed to fit and interpret the data. For example, in gravitational wave (GW) astrophysics, waveform models are constructed to match-filter the data. Such models are characterized by a set of physical parameters that describe the system that generated the GWs, e.g.~for NS binaries, these include the masses, the spin angular momenta, the quadrupole moments and the tidal deformabilities. If the EoS is known, one does not need to fit for the quadrupole moments and the tidal deformabilities, since these are functions of the masses and spins only. Lacking precise knowledge of the EoS, these quantities must be included in the model parameter list and fitted for, which then dilutes the accuracy to which other parameters can be measured.      

The aforementioned problem can be alleviated if EoS-independent relations between the moment of inertia ($I$), the tidal Love number and the quadrupole moment ($M_2$ or $Q$) of NSs can be found: enter the \emph{I-Love-Q} relations~\cite{I-Love-Q-Science,I-Love-Q-PRD}. Although these relations are not exactly EoS-independent, they are approximately so, with variations only at the percent level. After their initial discovery, the I-Love-Q relations were confirmed using different EoSs~\cite{lattimer-lim}, using binary systems~\cite{maselli}, using magnetized NSs~\cite{I-Love-Q-B} and proto-NSs~\cite{Martinon:2014uua}, allowing for rapid rotation~\cite{Pappas:2013naa,Chakrabarti:2013tca,Yagi:2014bxa,Stein:2013ofa,Chatziioannou:2014tha}, through a post-Minkowskian expansion~\cite{Chan:2014tva}, using NSs with anisotropic pressure~\cite{Yagi:2015hda} and in alternative theories of gravity~\cite{I-Love-Q-Science,I-Love-Q-PRD,Sham:2013cya,Doneva:2014faa,Pani:2014jra,Kleihaus:2014lba}. An important extension of the I-Love-Q relations was also recently found that relates all of the multipole moments of the exterior gravitational field of NSs to just the first three in an approximately EoS independent fashion~\cite{Yagi:2013sva,Yagi:2014bxa}. 

The I-Q relations may help in the independent determination of the NS mass and radius \cite{Baubock:2013gna,Psaltis:2013zja,Psaltis:2013fha,Morsink:2007tv,Baubock:2012bj} from observations of the X-ray pulse profiles of millisecond pulsars with NICER~\cite{2012SPIE.8443E..13G} and LOFT~\cite{2012AAS...21924906R,2012SPIE.8443E..2DF}. In principle, the X-ray profile depends on the independent parameters ($M$, $R$, $I$, $M_2$, f, $\iota$), where $M$ is the mass, $R$ is the radius, $f$ is the spin frequency and $\iota$ is the inclination angle. The I-Q relation can be used to eliminate $M_{2}$ in favor of $I$, and for stars with compactnesses $C \equiv M/R \in (0.1,0.4)$, the I-C relation is also approximately EoS independent. Such a reduction in the model parameter space may allow one to extract the mass and radius independently from X-ray observations with perhaps a $5\%$ accuracy~\cite{Baubock:2013gna,Psaltis:2013fha}.

Given these accuracy requirements, one must ensure that systematic errors are under control. One such source of error is in the EoS variability of the I-Q relations at the percent level. Another is in the effect of higher-order multipole moments, like the octopole moment $S_{3}$ and the hexadecapole moment $M_{4}$. These moments do not need to be included as new parameters in the pulse profile model because of the $S_{3}$-$M_{2}$ and the $M_{4}$-$M_{2}$ relations discovered in~\cite{Yagi:2013sva,Yagi:2014bxa}, which are also approximately EoS universal. The EoS universality in the latter, however, is weaker than in the I-Q relations, with EoS variability at the $15\%$ level, which may introduce systematic errors that could contaminate the extraction of the mass and radius~\cite{Yagi:2014bxa}. 

The EoS variability of the I-Q relations and the relations between higher multipole moments depends sensitively on how they are adimensionalized. This is related to the point made by~\cite{Pappas:2013naa} and~\cite{Chakrabarti:2013tca}, who discovered that the I-Q relations are approximately universal for rapidly rotating NSs \emph{provided} one fixes a \emph{dimensionless} spin parameter as one explores a sequence of NSs. If one fixes the dimensional spin frequency, the EoS variability greatly increases, as discovered in~\cite{doneva-rapid}. This, of course, does not mean that the EoS universality is lost, but rather that an inappropriate spin parameter was fixed in the sequence of stars. Such findings provide clear evidence that the EoS variability is sensitive to the parameters chosen to construct the relations.

Can one then find a better set of normalization constants that further reduces the EoS variability in the I-Q, $M_{4}$-$M_{2}$ and $S_{3}$-$M_{2}$ relations so that they can be used in the modeling of the X-ray pulse profile? The answer to this question is yes and it is the main topic of this paper. When the I-Love-Q relations were discovered~\cite{I-Love-Q-PRD}, the moment of inertia was adimensionalized via $\bar{I} \equiv I/M^3$, the quadrupole moment via $\bar{M_2} \equiv -M_2/(M^3\chi^2)$, the octopole moment via $\bar{S}_{3} = - S_{3}/(M^{4} \chi^{3})$ and the hexadecapole moment via $\bar{M}_{4} = M_{4}/(M^{5} \chi^{4})$, where $\chi\equiv S_1/M^2$ is a dimensionless spin-parameter, $S_{1} = I \Omega$ is the spin angular momentum, and $\Omega$ is the spin angular frequency. With such choices, the I-Q, $M_{4}$-$M_{2}$ and $S_{3}$-$M_{2}$ relations have an EoS variability of $2\%$, $15\%$ and $8\%$ respectively~\cite{Yagi:2014bxa} for slowly- or rapidly-rotating NSs and realistic EoSs~\cite{APR,SLy,shibata-fitting,LS,ott-EOS,Shen1,Shen2,Wiringa:1988tp,Alford:2004pf}. If instead of this choice, one normalizes the multipole moments via $\bar{I}^{\rm{new}} \equiv I/(M^3 C^{1.2})$, $\bar{M_2}^{\rm{new}} \equiv -M_2/(M^3\chi^2 C)$, $\bar{S}_{3}^{\rm{new}} = S_{3}/(M^{4} \chi^{3} C^{0.7})$ and $\bar{M}_{4}^{\rm{new}} = M_{4}/(M^{5} \chi^{4} C^{0.6})$, then the maximum EoS variability in the new I-Q, $M_{4}$-$M_{2}$ and $S_{3}$-$M_{2}$ decreases by a factor of $2$ or $3$, down to $\sim 1\%$, $\sim 6\%$, and $\sim 2\%$ respectively, for slowly-rotating NSs with the same EoSs. 

\begin{figure*}[htb]
\begin{center}
\includegraphics[width=8.5cm,clip=true]{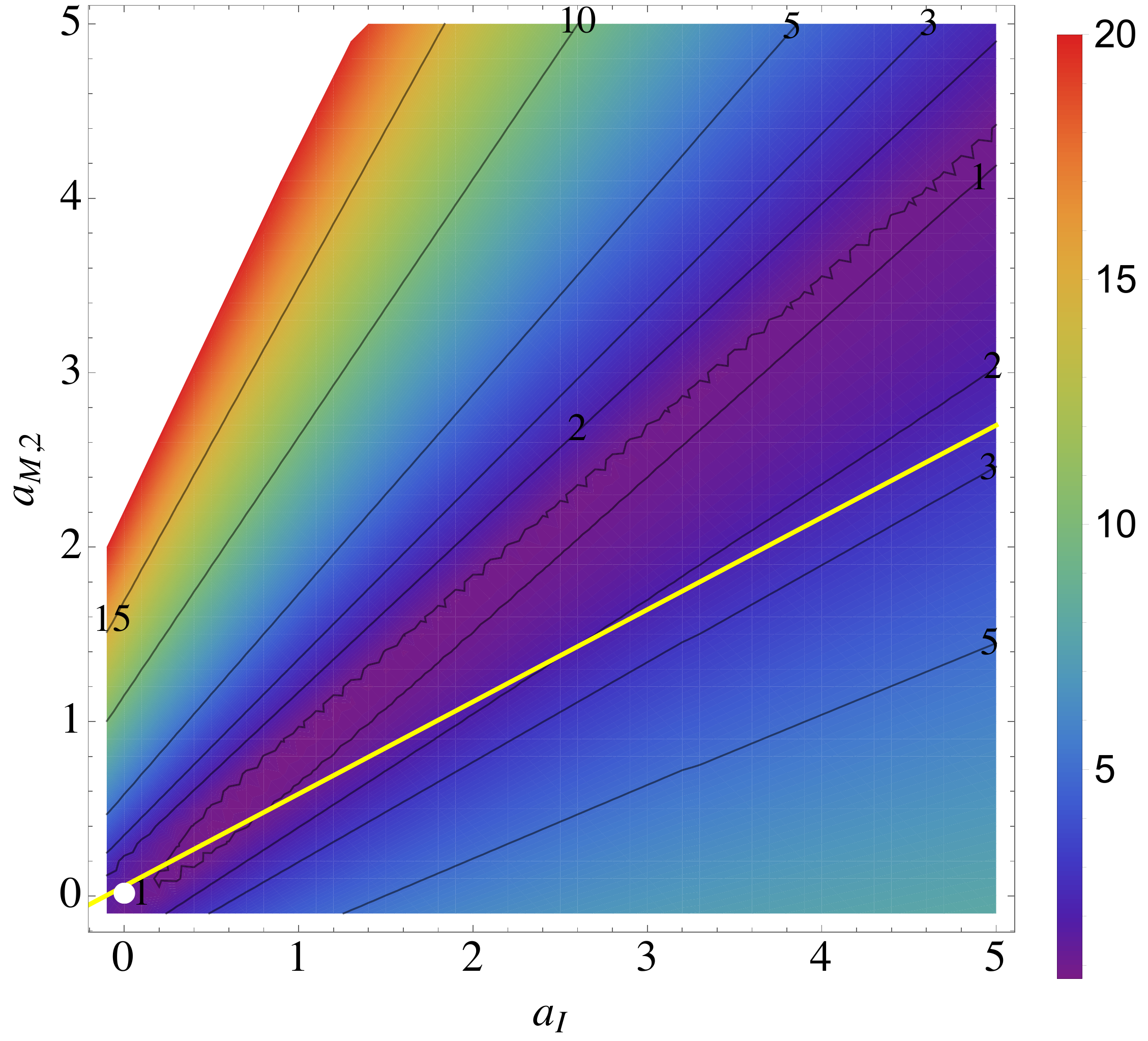}  \qquad
\includegraphics[width=8.5cm,clip=true]{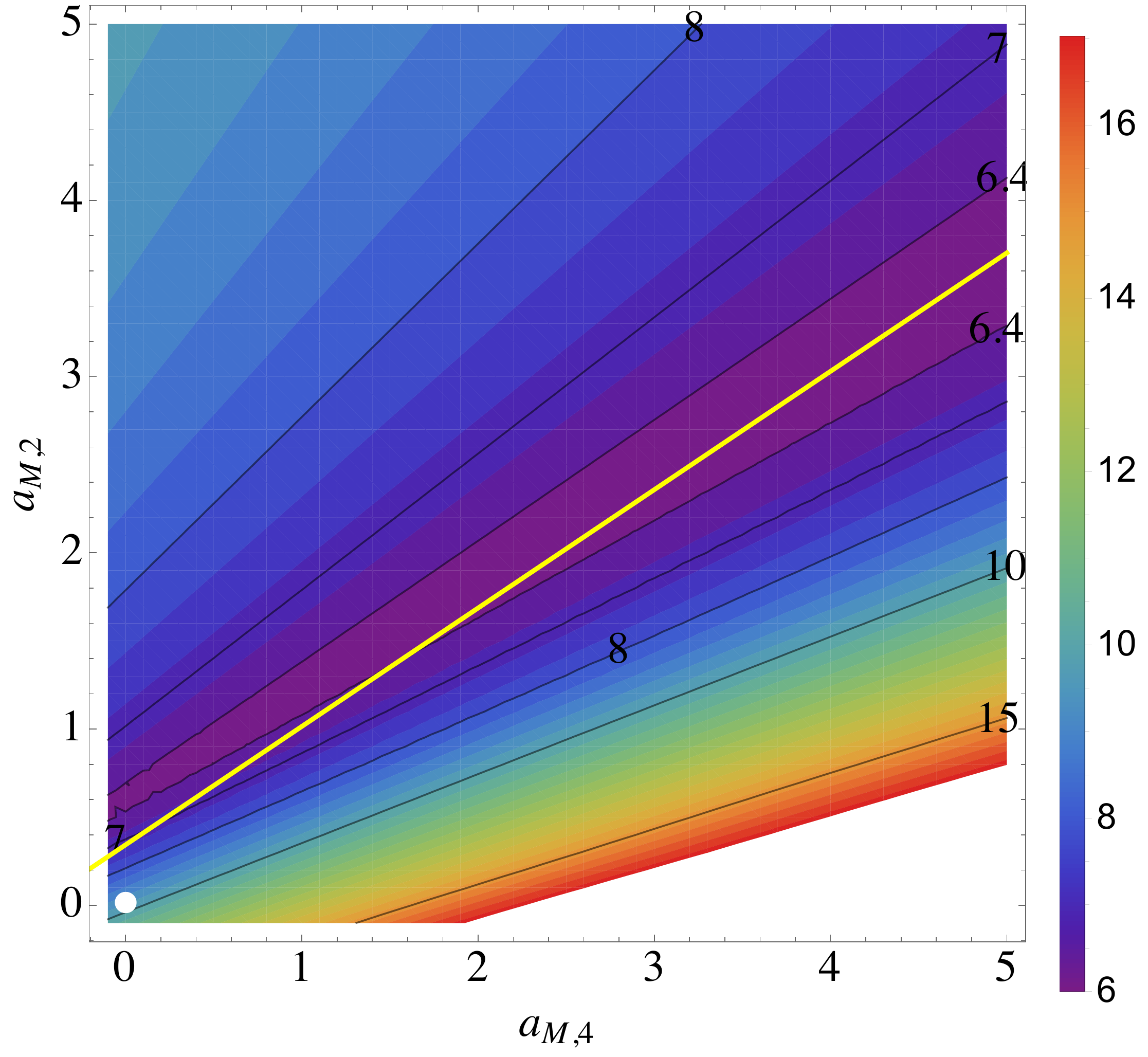}  
\caption{\label{fig:CP_IQ_half} (Color Online) EoS variability in the I-Q (left) and $M_{4}$-$M_{2}$ relations (right) for different values of the normalization constants $(a_{I},a_{M,2})$ (left) and $(a_{M,4},a_{M,2})$ (right). The moment of inertia is normalized via $\bar{I}=I/(M^3 C^{a_I})$, the quadrupole moment via $\bar{M}_2=-M_{2}/(M^3 \chi^2 C^{a_{M,2}})$, and the hexadecapole moment via $\bar{M}_4= {M_{4}}/({M^5 \chi^4 C^{a_{M,4}}})$. For a given choice of normalization, the contours show the maximum percent EoS variability in the I-Q and $M_{4}$-$M_{2}$ relations computed with variety of realistic EoSs~\cite{APR,SLy,shibata-fitting,LS,ott-EOS,Shen1,Shen2,Wiringa:1988tp,Alford:2004pf}, relative to a representative realistic EoS~\cite{LS}. For comparison, we also present the set of $(a_{I},a_{M,2})$ and $(a_{M,4},a_{M,2})$ that leads to the least EoS variability in a non-relativistic analysis (solid yellow line) for polytropic EoSs, as well as the original normalizations chosen in~\cite{I-Love-Q-Science,I-Love-Q-PRD,Yagi:2014bxa} (white solid circle). Observe that even in the fully relativistic calculation, the set of $(a_{I},a_{M,2})$ and $(a_{M,4},a_{M,2})$ that minimizes the EoS variability is a straight line in this space, where the slope and $y$-intercepts are different from that obtained with a Newtonian analysis. Observe also that there are choices of normalization that lead to less EoS variability than the original normalization of~\cite{I-Love-Q-Science,I-Love-Q-PRD,Yagi:2014bxa}.}
\end{center}
\end{figure*}

The above choices of normalization are not unique, and in fact, there is an entire family of normalizations that minimizes the degree of EoS variability. To find this family, we divide our study in two parts: (i) an analytic, non-relativistic treatment, where we consider polytropic EoS with index $n \in (0,1)$; and (ii) a numerical, fully relativistic analysis where we consider realistic EoS~\cite{APR,SLy,shibata-fitting,LS,ott-EOS,Shen1,Shen2,Wiringa:1988tp,Alford:2004pf}. In both cases, we focus on slowly-rotating stars with masses $M \in (1,2.5) M_{\odot}$ in the Hartle-Thorne formalism~\cite{hartle1967,hartlethorne} and without magnetic fields, as this is appropriate for recycled pulsars, even with millisecond spin periods. We compute the multipole moments for these stars and then adimensionalize them in the same way as originally discovered in~\cite{I-Love-Q-PRD} and explained above, but with an extra factor of $C^{a_{p}}$, where the power $a_{p}$ is different for each moment. Each of the I-Q, $M_{4}$-$M_{2}$ and $S_{3}$-$M_{2}$ relations then depends on two free parameters, $(a_{I},a_{M,2})$, $(a_{M,4},a_{M,2})$ and $(a_{S,3},a_{M,2})$. We discretize this space and compute the maximum EoS variability at each point to find the set that minimizes it. In both cases, this set can be described by a straight line in $(a_{I},a_{M,2})$, $(a_{M,4},a_{M,2})$ and $(a_{S,3},a_{M,2})$ space, as shown in Fig.~\ref{fig:CP_IQ_half}. The slope and $y$-intercept of this line is slightly different in the non-relativistic and in the relativistic calculations. Observe that the best normalization does not always agree with that originally chosen in~\cite{I-Love-Q-PRD}. Similar results are found for the $S_{3}$-$M_{2}$ relation. 

The remainder of this paper deals with the details of the results discussed above and it is organized as follows. 
In Sec.~\ref{sec:original}, we review the original I-Q and three-hair relations for NSs, as derived in~\cite{Stein:2013ofa}. 
In Sec.~\ref{sec:Newton}, we re-formulate the equations in the non-relativistic limit with a generic normalization to study how the EoS universality depends on this. 
In Sec.~\ref{sec:GR}, we repeat this analysis but in full General Relativity. 
In Sec.~\ref{sec:conclusion} we conclude and highlight possible directions for future research. 
All throughout, we use geometric units in which $G=1= c$.

\section{Original Three-Hair Relations}
\label{sec:original}

In this section, we briefly review the three-hair relations for NSs derived in~\cite{Stein:2013ofa}. The exterior gravitational field of an isolated and stationary mass distribution can be written in terms of a multipole moment decomposition~\cite{MTW}. In the non-relativistic limit (i.e.~in the perturbative weak field, when we expand all expressions to leading order in powers of compactness) and in a slow rotation expansion (i.e.~an expansion in powers of the product of the mass and the spin angular frequency), the leading-order expressions for the mass and mass-current multipole moments are~\cite{Ryan:1996nk}
\begin{align}
\label{m01}
M_{\ell} = 2\pi \int_0^{\pi} \int_0^{R_*(\theta)} \!\!\! \rho(r,\theta) P_{\ell}(\cos \theta) \sin \theta d\theta \; r^{\ell+2} \; dr~~,
\end{align}
and 
\begin{align}
\label{s02}
S_{\ell} = \frac{4\pi \Omega}{\ell+1} \int_0^{\pi} \int_0^{R_*(\theta)} \!\! \rho(r,\theta) \frac{dP_{\ell}(\cos \theta)}{d\cos \theta} \sin^3 \theta d\theta \; r^{\ell+3} \; dr ~~,
\end{align}
where $\rho(r,\theta)$ is the density, $R_*(\theta)$ is the stellar surface profile, $\Omega$ is the spin angular velocity and $P_{\ell}(\cos \theta)$ are the Legendre polynomials of order $\ell$. These moments are distinct but algebraically related to the Geroch-Hansen moments~\cite{Geroch:1970cc,Geroch:1970cd,hansen:46} and the Thorne moments~\cite{Thorne:1980rm} in the non-relativistic limit~\cite{pappas-apostolatos,Pappas:2013naa}. 

In the non-relativistic limit, slowly-rotating NSs can be well modeled in the \emph{elliptical isodensity approximation} of~\cite{Lai:1993ve}. In this scheme, one assumes that isodensity surfaces are self-similar ellipsoids and that the density profile is the same as that of a non-rotating star of the same volume as the rotating star. Within this approximation, and in a suitable coordinate system, the angular and radial integrals of Eqs.~\eqref{m01} and~\eqref{s02} can be separated in terms of certain angular integrals $I_{\ell,\ell'}$ and certain radial integrals $R_{\ell}$, namely
\begin{equation}
\label{m04}
M_{\ell} = 2\pi I_{\ell,3} R_{\ell} ~~,
\end{equation}
and 
\begin{equation}
\label{s05}
S_{\ell} = \frac{4\pi \ell}{2\ell+1}\Omega (I_{\ell-1,5}-I_{\ell+1,3})R_{\ell+1}~~.
\end{equation}
The angular integrals can be calculated in closed form and are~\cite{2007tisp.book.....G}
\begin{equation}
I_{\ell,3} = (-1)^{\ell/2}\frac{2}{\ell+1}\sqrt{1-e^2}e^{\ell}~~,
\end{equation}
and
\begin{equation}
I_{\ell-1,5} - I_{\ell+1,3} = (-1)^{(\ell-1)/2}\frac{2(2\ell+1)}{\ell(\ell+2)}\sqrt{1-e^2}e^{\ell-1}~~.
\end{equation}
The radial integrals can be rewritten using the Lane-Emden function~\cite{2004sipp.book.....H} as
\begin{equation}
R_{\ell} = \rho_c \left( \frac{a_1}{\xi_1} \right)^{\ell+3}{\cal R}_{n,\ell}~~,
\end{equation}
where $\theta = (\rho/\rho_{c})^{1/n}$ is the Lane-Embden function, $\rho_{c}$ is the central density and 
$n$ is a constant, later to be identified with the polytropic index. The reduced radial integral is defined as
\begin{equation}
{\cal R}_{n,\ell} = \int_0^{\xi_1} [\vartheta(\xi)]^n \xi^{\ell+2}d\xi~~,
\end{equation}
where $M (=M_0)$ is the stellar mass of the non-rotating configuration, $\xi=(\xi_1/a_1)\tilde{r}$ is the dimensionless radial coordinate, with $a_{1}$ the semi-major axis of the ellipse, and $\xi=\xi_1$ is the dimensionless stellar radius. Observe that all of the EoS dependence is here encoded in the $\Omega(e)$ function and the radial integrals $R_{\ell}$. Notice also that all the results reviewed above assume a single polytropic EoS, i.e.~an EoS of the form $p = K \rho^{1 + 1/n}$, but they can be easily extended to piecewise polytropes~\cite{Chatziioannou:2014tha}, which accurately approximate realistic EoSs ~\cite{read-markakis-shibata,lattimer-prakash-review}. 

With all this information at hand, the mass and mass-current moments can be written as
\begin{equation}
\label{m10}
M_{2\ell+2}=\frac{(-1)^{\ell+1} e^{2 \ell+2}  M^{2 \ell+3} {\cal R}_{n,2 \ell+2}}{(2 \ell+3)~\left(1-e^2\right)^{\frac{\ell+1}{3}}~\xi_1^{2 \ell+4}~C^{2 \ell+2}~\vert \vartheta'(\xi_1)\vert}
\end{equation}
and
\begin{equation}
\label{s11}
S_{2\ell+1}=\frac{(-1)^{\ell} ~2~\Omega(e)~ e^{2 \ell}~  M^{2 \ell+3} {\cal R}_{n,2 \ell+2}}{(2 \ell+3)~\left(1-e^2\right)^{\frac{\ell+1}{3}}~\xi_1^{2 \ell+4}~C^{2 \ell+2}~\vert \vartheta'(\xi_1)\vert}\,,
\end{equation}
where $C = M/R$ is the stellar compactness. In the elliptical isodensity approximation, the angular frequency is simply~\cite{Lai:1993ve}
\begin{equation}
\label{o12}
\Omega(e)=\frac{3}{2}\xi_1^2 \left[\frac{C^3\vert \vartheta'(\xi_1)\vert}{(5-n)M^2~{\cal R}_{n,2}}\right]^{1/2}~f(e) ~~,
\end{equation}
where $f(e)=[-6e^{-2}(1-e^2)+2e^{-3}(1-e^2)^{1/2}(3-2e^2)\arcsin(e)]^{1/2}$. 

Let us now derive the three-hair relations. First, we adimensionalize Eqs.~\eqref{m10} and~\eqref{s11} via 
\begin{equation}
\bar{M}_{\ell} = (-1)^{\ell/2}\frac{M_{\ell}}{M^{\ell+1}\chi^{\ell}}~~,~~~~\bar{S}_{\ell}=(-1)^{\frac{\ell-1}{2}}\frac{S_{\ell}}{M^{\ell+1}\chi^{\ell}}~~,
\label{eq:old-norm}
\end{equation}
where $\chi=S_1/M^2$ is the dimensionless spin parameter. Then, we set $\ell=0$ in the adimensionalized $\bar{M}_{2\ell+2}$ equation to solve for compactness $C$ as a function of $\bar{M}_{2}$. And finally, we substitute this expression back both into the adimensionalized version of Eq.~\eqref{m10} and Eq.~\eqref{s11} to obtain $\bar{M}_{2\ell+2}$ and $\bar{S}_{2 \ell+1}$ as functions of $\bar{M}_{2}$. These expressions can then be expressed via the single equation
\begin{equation}
\bar{M}_{2\ell+2} + i\bar{S}_{2\ell+1} = \bar{B}_{n,\ell}\bar{M}_2^{\ell}(\bar{M}_2 + i\bar{S}_1) ~~,
\end{equation}
where 
\be
\bar{B}_{n,\ell}=\left[
\frac{(2\ell+3)^{1/\ell}}{3^{1+1/\ell}}\frac{{\cal R}_{n,2}^{1+1/\ell} {\cal R}_{n,2\ell+2}^{-1/\ell}}{\vert \vartheta'(\xi_1)\vert \xi_1^2}
\right]^{-\ell}
\ee
This is the so-called \emph{three-hair relation} for NSs in the non-relativistic limit, which is approximately EoS independent. Indeed, all of the EoS dependence is encoded in the $\bar{B}_{n,\ell}$ coefficients, which were shown to vary weakly with $n$ for any given $\ell$~\cite{Stein:2013ofa}. Similar expressions hold in the relativistic regime for rapidly rotating stars~\cite{Pappas:2013naa,Chakrabarti:2013tca,Yagi:2014bxa}.  

The I-Q relations can be derived from the expressions presented above. Setting $\ell=0$ in Eqs.~\eqref{m10}--\eqref{o12}, we then obtain
\begin{equation}
\label{m2ieq1}
\bar{M}_2 = \frac{\bar{I}e^2}{2\chi^2} ~~.
\end{equation}
This equation, however, is not yet the I-Q relation we want, because the right-hand side still depends 
on  $e$ and $\chi$. We can solve for $\chi$ as a function of $e$ by setting $\ell=0$ in Eq.~\eqref{s11}, 
solving for compactness and using the result in the definition of $\chi = S_{1}/M^{2}$. Doing so, we find
\begin{equation}
\label{chieqn}
\chi = \frac{\sqrt{15}}{4}\frac{\bar{I}^{1/4}}{(1-e^2)^{1/4}A_{n,0}^{1/2}}f(e)\,,
\end{equation}
and substituting this into Eq.~\eqref{m2ieq1} we obtain the I-Q relation. 
Expanding this in the slow-rotation limit, we can write the I-Q relation as
 \begin{equation}
\label{im2rel}
\bar{M}_2 = \sqrt{\bar{I}} A_{n,0}\,,
\end{equation}
where
\begin{equation}
\label{a21}
A_{n,0}=\left\{\left[\frac{25(5-n)^2}{1152}\right]\frac{27 \; {\cal R}_{n,2}^{2}}{{\cal R}_{n,2}^3 \xi_1^{-4} \vert \vartheta'(\xi_1)\vert^{-1}}\right\}^{\frac{1}{2}} ~~.
\end{equation} 

\section{Toward Exact Universality: \\ Analytical Results in the Non-Relativistic Limit}
\label{sec:Newton}

Can the approximate EoS universality of the NS no-hair relations be improved by modifying the normalization of the multipole moments? 
Let us then re-define the dimensionless multipole moments as
\begin{equation}
\label{Meqn}
\bar{M}_{2\ell+2}^{(a_{M,2\ell+2})}=\frac{(-1)^{\ell+1} M_{2 \ell+2}}{M^{2 \ell+3} \chi^{2 \ell+2} C^{a_{M,2 \ell+2}}}
\end{equation}
and
\begin{equation}
\label{Seqn}
\bar{S}_{2\ell+1}^{(a_{S,2\ell+1})}=\frac{(-1)^{\ell} S_{2 \ell+1}}{M^{2 \ell+2} \chi^{2 \ell+1} C^{a_{S,2 \ell+1}}}\,,
\end{equation}
where $a_{M,2 \ell+2}$ and $a_{S,2\ell+1}$ are real constants that provide flexibility in the adimensionalization. Notice that to recover the original normalization, one must set $a_{M,2 \ell + 2}=0$ and $a_{S,2 \ell + 1}= 0$.

The approximate no-hair relations can now be obtained with the new normalization. First, we eliminate the compactness from the dimensionless multipole moments by solving Eq.~\eqref{Meqn} with $\ell=0$ for $C$ in terms of $\bar{M}_2^{(a_{M,0})}$. Substituting this back into Eq.~\eqref{Meqn} and Eq.~\eqref{Seqn}, we find 
\begin{align}
\label{mbareqn}
\bar{M}_{2\ell+2}^{(a_{M,2\ell+2})} &=\frac{4^{\alpha}}{2 \ell+3}~3^{-\alpha+\ell+1}~\left(\frac{M\Omega}{e}\right)^{2\alpha}~(1-e^2)^{-\alpha/3} \nonumber \\
& \times \vert \vartheta'(\xi_1)\vert^{\ell-\alpha}~\xi_1^{2(\ell-2\alpha)}~{\cal R}_{n,2}^{\alpha-\ell-1}~{\cal R}_{n,2 \ell+2} \nonumber \\
& \times \left(\bar{M}_2^{(a_{M,0})}\right)^{\alpha+\ell+1}~~,
\\
\label{sbareqn}
\bar{S}_{2\ell+1}^{(a_{S,2\ell+1})} &=\frac{4^{\beta}}{2 \ell+3}~3^{-\beta+\ell+1}~\left(\frac{M\Omega}{e}\right)^{2\beta}~(1-e^2)^{-\beta/3} \nonumber \\
& \times \vert \vartheta'(\xi_1)\vert^{\ell-\beta}~\xi_1^{2(\ell-2\beta)}~{\cal R}_{n,2}^{\beta-\ell-1}~{\cal R}_{n,2 \ell+2} \nonumber \\
& \times \left(\bar{M}_2^{(a_{M,0})}\right)^{\beta+\ell}~~,
\end{align}
where $\alpha$ and $\beta$ are new constants defined by 
\begin{align}
\alpha &= \frac{(\ell+1)a_{M,2}-a_{M,2\ell+2}}{-a_{M,2}+2}\,,
\\
\beta &= \frac{\ell a_{M,2}+ -a_{S,2\ell+1}}{-a_{M,2}+2}\,.
\end{align}

Unlike in the original no-hair relations, summarized in Sec.~\ref{sec:original}, this time the multipole moments depend on the spin angular frequency $\Omega$ and the eccentricity $e$. One can eliminate this dependence by choosing $a_{M,2 \ell+2}= (\ell+1) a_{M,2}$ and $a_{S,2 \ell+1}= \ell a_{M,2}$, which leads to $\alpha = 0 = \beta$, and thus to 
\begin{equation}
\bar{M}_{2\ell+2}^{(a_{M,2\ell+2})}=\frac{3^{\ell+1} ~\vert \vartheta'(\xi_1)\vert^\ell ~\xi_1^{2 \ell} ~\bar{M}_2^{\ell+1}~ {\cal R}_{n,2 \ell+2}}{(2 \ell+3)~{\cal R}_{n,2}^{\ell+1}} ~~.
\end{equation}
and
\begin{equation}
\bar{S}_{2\ell+1}^{(a_{S,2\ell+1})}=\frac{3^{\ell+1} ~\vert \vartheta'(\xi_1)\vert^\ell ~\xi_1^{2 \ell} ~\bar{M}_2^{\ell}~ {\cal R}_{n,2 \ell+2}}{(2 \ell+3)~{\cal R}_{n,2}^{\ell+1}} ~~.
\end{equation}
These two expressions are similar to Eqs. (10) and (11) of~\cite{Stein:2013ofa}, but here, they are obtained for a larger class of normalizations, that in particular, includes the original normalization of~\cite{Stein:2013ofa}. 

With these general expressions at hand, let us now investigate a few examples of universality, beginning with the I-Q relations. Generalizing Eq. (22) of~\cite{Stein:2013ofa}, we adimensionalize the moment of inertia via 
\begin{equation}
\label{eq:I-Norm}
\bar{I}^{(a_I)}=\frac{\bar{I}}{C^{a_I}}=\frac{S_1}{\Omega M^3 C^{a_I}} ~~,
\end{equation}
where $a_I$ is a new normalization constant. Proceeding in the exact same way as in the previous section, the expression for $\chi$, still defined via $\chi \equiv S_{1}/M^{2}$, as a function of eccentricity is
\begin{equation}
\label{chidef}
\chi = \frac{\sqrt{3}(1-e^2)^{-\frac{1}{9}-\frac{x}{3}} \left(\bar{I}^{(a_I)}\right)^{\frac{2}{3}-x}(5-n)^{-\frac{5}{6}+2x}}{2^{-\frac{1}{2}+6x}5^{\frac{1}{3}-2x}A_{n,0}^{-\frac{1}{3}+2x}}~f(e)~~,
\end{equation}
where $x=\frac{5+4a_I}{6(2+a_I)}$and $A_{n,\ell}$ is still given by Eq.~\eqref{a21}. Clearly, we recover Eq.~\eqref{chieqn} when $a_I=0$, and thus $x=5/12$. Solving for compactness from the $\ell=0$ version of Eq.~\eqref{mbareqn}, and using Eq.~\eqref{chidef}, we get the I-Q relation in the slow rotation limit:
\begin{align}
\label{eq:new-IQ}
\bar{M}_2^{(a_{M,2})} &= \left(\frac{2}{3}\right)^{\frac{a_I-2 a_{M,2}}{2 a_I+4}}~\vert \vartheta'(\xi_1)\vert^{-\frac{a_I-2 a_{M,2}}{2 a_I+4}}~\xi_1^{-\frac{2 \left(a_I-2 a_{M,2}\right)}{a_I+2}} \nonumber \\
& {\cal R}_{n,2}^{\frac{a_I-2 a_{M,2}}{2 a_I+4}}~A_{n,0}~\bar{I}^{\frac{a_{M,2}+1}{a_I+2}}~~.
\end{align}
Notice that the new I-Q relation depends only on the set $(a_{M,2},a_{I})$. Clearly, we recover Eq.~\eqref{im2rel} when $a_{M,2}=a_I/2$, which includes the original normalization of~\cite{Stein:2013ofa}.
 
Given the new I-Q relation in Eq.~\eqref{eq:new-IQ}, we can now address whether there are particular sets of parameters $(a_{M,2},a_{I})$ for which the EoS universality is improved, i.e.~for which the variability of the I-Q relation with different EoSs is decreased. One way to assess this variability is to compute the relative fractional difference in the I-Q relation between an $n=0$ and an $n=1$ polytropic EoS. This is analytically tractable because exact solutions to the equations of stellar structure exist for such polytropic EoSs. This fractional difference reduces to 
\begin{align}
\label{frdin0n1}
& \frac{\bar{M}_2^{(a_{M,2})}\vert_{n=0}-\bar{M}_2^{(a_{M,2})}\vert_{n=1}}{\bar{M}_2^{(a_{M,2})}\vert_{n=0}} = \nonumber \\
& 1-4 \cdot 5^{-\frac{a_{M,2}+a_I+3}{a_I+2}} \pi^{\frac{2 \left(a_{M,2}+1\right)}{a_I+2}} \left(\frac{3}{\pi^2-6}\right)^{\frac{a_{M,2}+1}{a_I+2}}\,,
\end{align}
and setting it to zero, we can determine the parameters $(a_{M,2},a_{I})$ that lead to exact universality between these two EoSs. 

\begin{figure}[t]
\begin{center}
\includegraphics[width=8.5cm,clip=true]{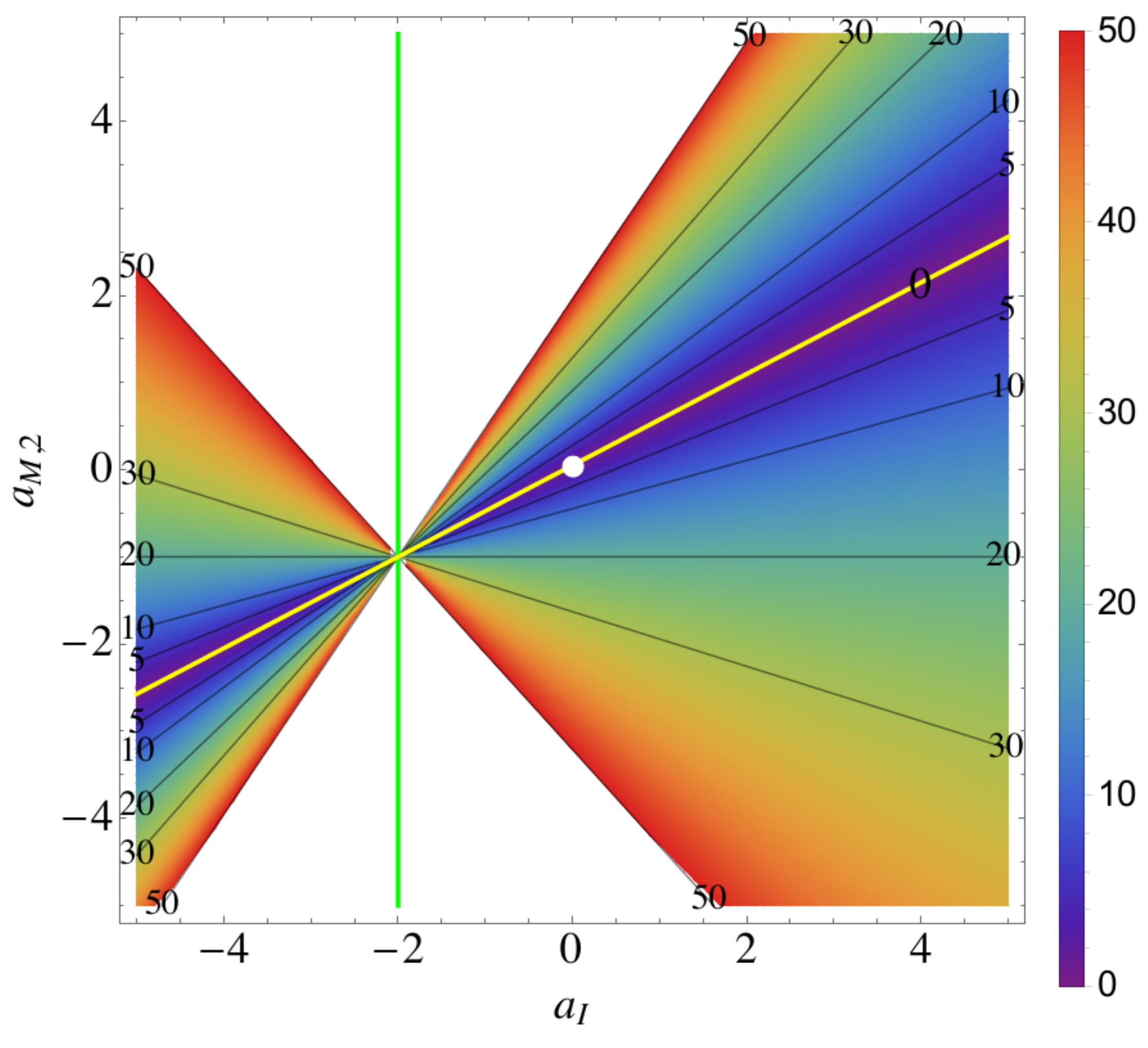}  
\caption{\label{fig:fracdiffn0n1} (Color Online) Relative fractional difference in the Newtonian I-Q relations for different $(a_{M,2},a_{I})$ normalizations. The fractional difference is here calculated by computing the I-Q relations with various polytropic EoSs relative to a polytropic EoS with index $n = 0.643$. The set of $(a_{M,2},a_{I})$ in the violet region leads to the most EoS independence in the I-Q relations, while the set in the red region leads to a variation of the relations with EoS of $\sim$ 50\%.  The green line shows the choices of $(a_{M,2},a_{I})$ for which the relative fractional difference is not well-defined. The white regions lead to EoS variability in the I-Q relations that exceeds 50\%. The yellow line shows the one-parameter family of $(a_{M,2},a_{I})$ that leads to zero relative fractional difference in the I-Q relations when computed with an $n=0$ and an $n=1$ polytropic EoS. The original normalization of~\cite{I-Love-Q-PRD,I-Love-Q-Science} is shown with a white circle. Observe that this normalization lies very close to the optimal one-parameter family.}
\end{center}
\end{figure}

Figure~\ref{fig:fracdiffn0n1} presents contours of the relative fractional difference, including the zero contour, in the $(a_{M,2},a_{I})$ plane.  Observe that there exists a one-parameter family of $(a_{M,2},a_{I})$ for which the relative fractional difference is exactly zero. Observe also that the original normalization is very close to this one-parameter family.~\footnote{Although not easily seen in the figure, the original normalization does not actually lie on the line.} A third interesting point is that the universality deteriorates rapidly for values of the normalization constants away from the one-parameter family. Finally, observe that there are values of $(a_{M,2},a_{I})$ for which Eq.~\eqref{frdin0n1} diverges, e.g.~if you take the limit as $a_{I} \to -2$ from the left, Eq.~\eqref{frdin0n1} diverges to positive infinity. This is because in such a limit, $S_1$ does not depend on $C$, which in turn means that one cannot solve $C$ for $\bar{I}^{(-2)}$, which is crucial in deriving the I-Q relations.
%

\begin{figure}[t]
\begin{center}
\includegraphics[width=8.5cm,clip=true]{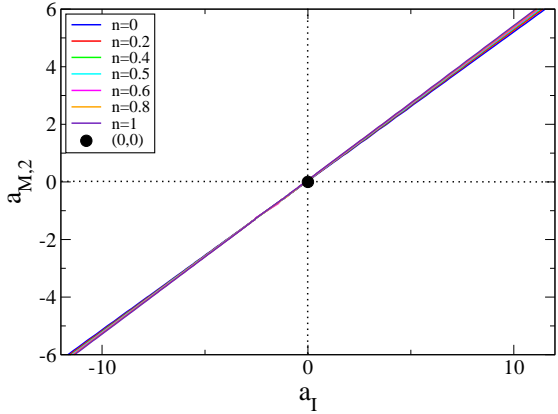}  
\caption{\label{fig:LEfrac} (Color Online) Values of $(a_{M,2},a_I)$ that lead to zero relative fractional difference in the I-Q relations calculated with polytropic EoSs with $n\in[0,1]$ and a reference polytropic EoS with $n=0.643$. The quadrupole moment and the moment of inertia are computed by numerically solving the Lane-Emden equation. Observe that all lines cluster around each other. Observe also that the original normalization [$(a_{M,2},a_I)=(0,0)]$ used in \cite{I-Love-Q-PRD,I-Love-Q-Science} lies very close to any of the lines.}
\end{center}
\end{figure}

These results suggest that one may find some choices of normalizations for which the I-Q relations are perfectly universal with respect to any EoS. To investigate this further, let us re-compute the relative fractional difference with polytropic EoSs of different indices. Given any polytropic index, we numerically solve the Lane-Emden equation and compute the relative fractional difference of the I-Q relations with that index and a \emph{reference} EoS. For the latter, we choose a polytropic EoS with index $n=0.643$, which is the average index of all of those presented in Table I of~\cite{Chatziioannou:2014tha}; the latter accurately approximate realistic EoSs~\cite{Read:2008iy} that support NSs with masses greater than $2 M_{\odot}$~\cite{2.01NS}. Figure~\ref{fig:LEfrac} shows the one-parameter family of $(a_{M,2},a_{I})$ for which this relative fractional difference is exactly zero. Observe that all the one-parameter families are clustered around a single line. This indicates that there exists an optimal set of ($a_{M,2},a_I$) for which the I-Q relation is \emph{almost} exactly independent of the EoS. Observe also that this optimal set happens to be very close to the $(a_{M,2},a_I) = (0,0)$ point, which corresponds to the original normalization of the I-Q relations~\cite{Yagi:2013sva,Stein:2013ofa,Yagi:2014bxa}. 

\begin{figure*}[htb]
\begin{center}
\includegraphics[width=8.0cm,clip=true]{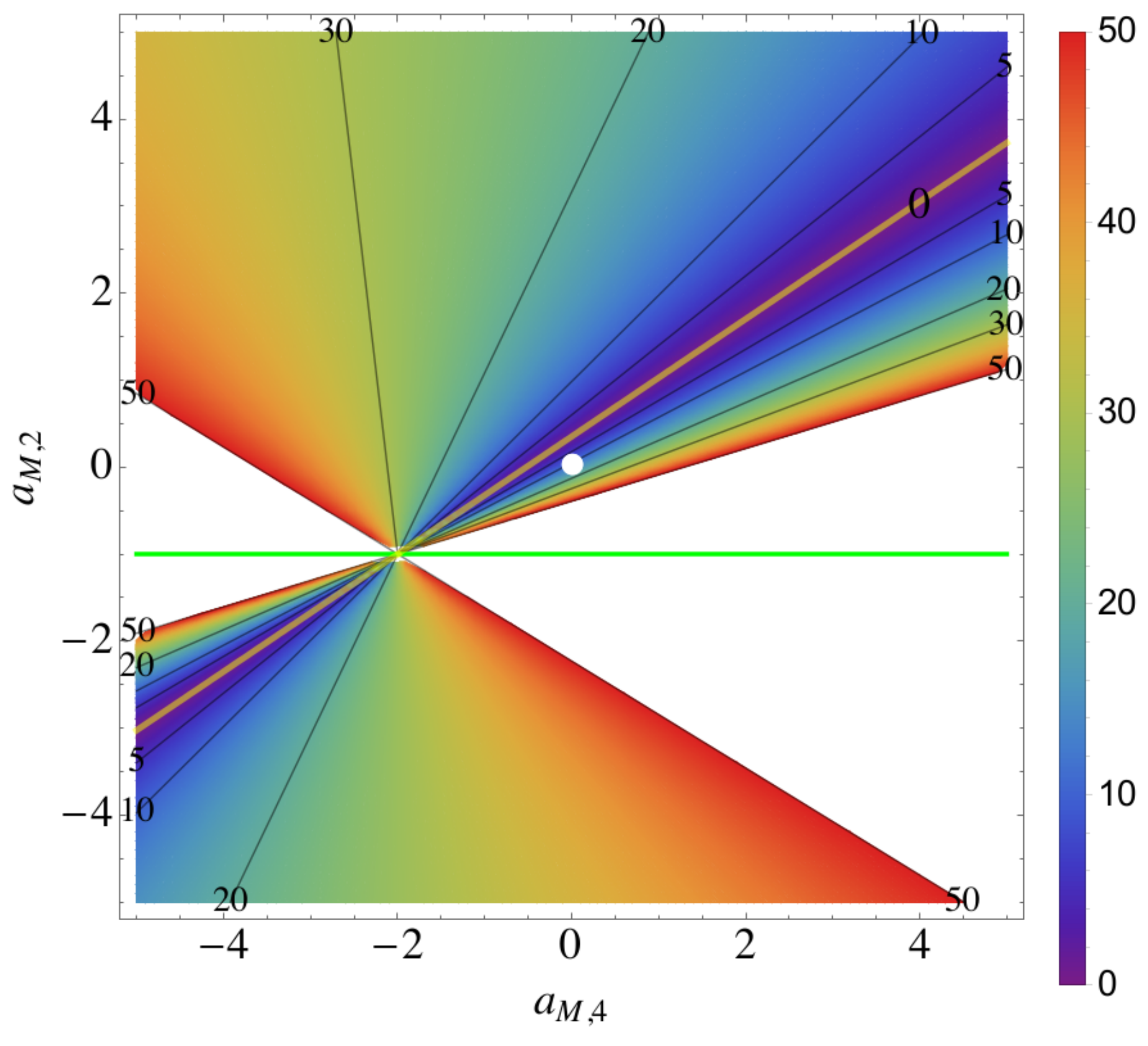} 
\hspace{0.5cm}
\includegraphics[width=8.0cm,clip=true]{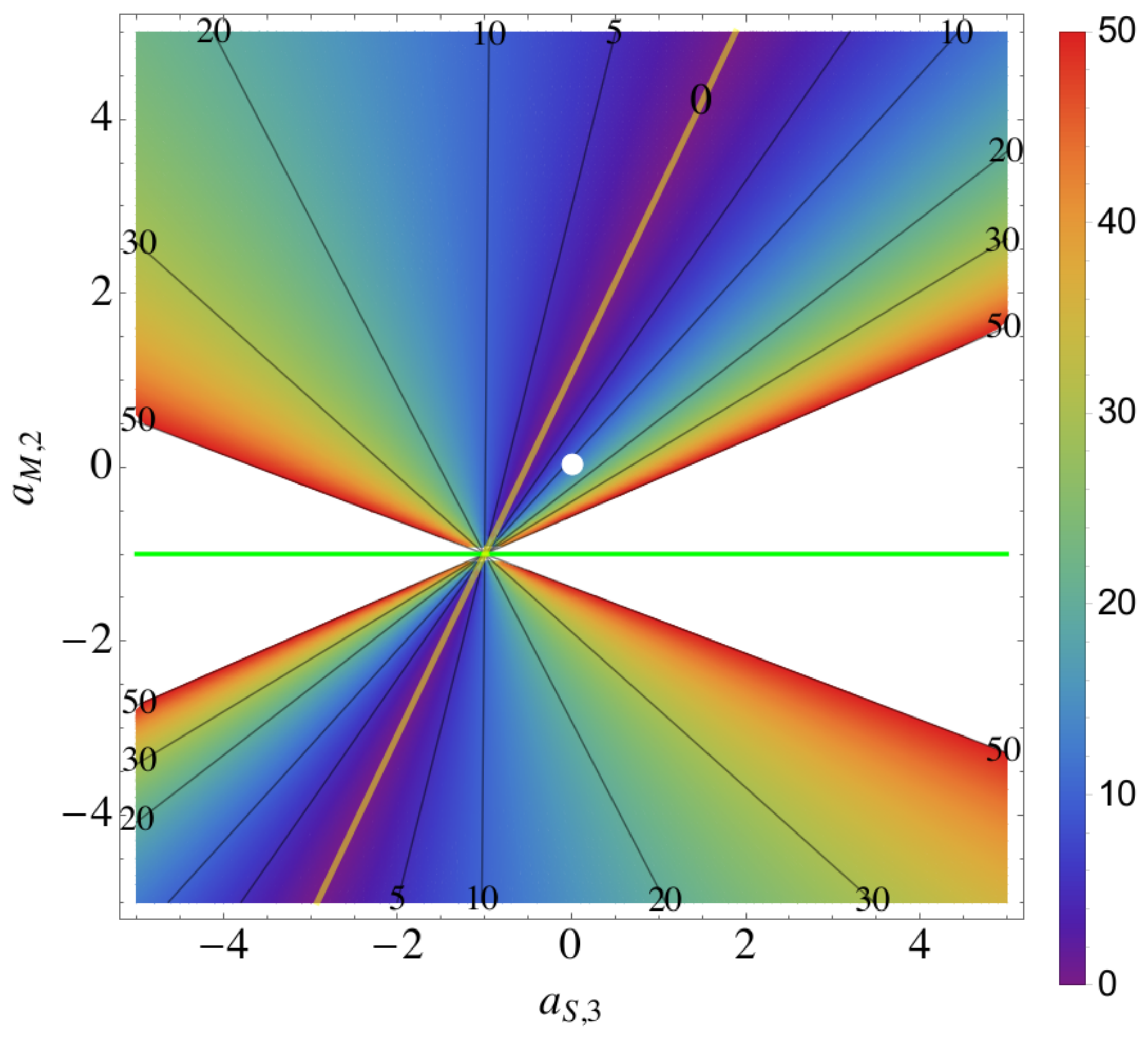}  
\caption{\label{fig:fracdiff2324n0n1} (Color Online) Same as Fig.~\ref{fig:fracdiffn0n1} but using the $M_4$-$M_2$ in the $(a_{M,2},a_{M,4})$ plane (left) and $S_{3}$-$M_{2}$ in the $(a_{S,3},a_{M,2})$ plane (right) relations. Observe that the original normalizations of~\cite{I-Love-Q-PRD,I-Love-Q-Science,Yagi:2014bxa} are close but do not quite coincide with the choices of normalization constant that minimizes the EoS variability in the relations.}
\end{center}
\end{figure*}

Let us now investigate another example of EoS universality as a function of the normalization constants, focusing on higher multipole moments. In particular, let us focus on the $\bar{M}_4-\bar{M}_2$ and $\bar{S}_3-\bar{M}_2$ relations, which  can be rewritten as
\begin{align}
\bar{M}_4^{(a_{M,4})} &= 9\times 2^{3(\delta_1-2)} 5^{1-\delta_1} (5-n)^{2-\delta_1} \nonumber \\
& \times \frac{\vert \vartheta'(\xi_1) \vert \xi_1^2 {\cal R}_{n,4} \bar{M}_2^{\delta_1}}{{\cal R}_{n,2}^2} ~~,
\end{align}
where $\delta_1=\frac{a_{M,4}+2}{a_{M,2}+1}$, and 
\begin{align}
\bar{S}_3^{(a_{S,3})} &= 9\times 2^{3(\delta_2-1)} 5^{-\delta_2} (5-n)^{1-\delta_2} \nonumber \\
& \times \frac{\vert \vartheta'(\xi_1) \vert \xi_1^2 {\cal R}_{n,4} \bar{M}_2^{\delta_2}}{{\cal R}_{n,2}^2} ~~,
\end{align}
where $\delta_2=\frac{a_{S,3}+1}{a_{M,2}+1}$, using Eqs.~\eqref{mbareqn} and \eqref{sbareqn} with $\ell=1$. As before, we compute the relative fractional difference in these relations between an $n=0$ and an $n=1$ polytropic EoS, since exact solutions to the stellar structure equations exist in these cases:
\begin{align}
\label{frac24eqn}
& \frac{\bar{M}_4^{(a_{M,4})}\vert_{n=0}-\bar{M}_4^{(a_{M,4})}\vert_{n=1}}{\bar{M}_4^{(a_{M,4})}\vert_{n=0}} = \frac{2^{-2\delta_1}}{625 \left(\pi ^2-6\right)^2} \nonumber \\
& \times \bigg[ \pi ^4 \left(625\times 2^{2\delta_1}-336\times 5^{\delta_1}\right) \nonumber \\
& -3 \pi ^2 \left(625\times 2^{2(\delta_1+1)}-448\times 5^{\delta_1+1}\right) \nonumber \\
& +9 \left(625\times 2^{2(\delta_1+1)}-896\times 5^{\delta_1+1}\right) \bigg]
\end{align}
and 
\begin{align}
\label{frac23eqn}
& \frac{\bar{S}_3^{(a_{S,3})}\vert_{n=0}-\bar{S}_3^{(a_{S,3})}\vert_{n=1}}{\bar{S}_3^{(a_{S,3})}\vert_{n=0}} = \frac{2^{-2\delta_2}}{125 \left(\pi ^2-6\right)^2} \nonumber \\
& \times \bigg[ \pi ^4 \left(125\times 2^{2\delta_2}-84\times 5^{\delta_2}\right) \nonumber \\
& +\pi ^2 \left(336\times 5^{\delta_2+1}-375\times 2^{2(\delta_2+1)}\right) \nonumber \\
& +9 \left(125\times 2^{2(\delta_2+1)}-224\times 5^{\delta_2+1}\right) \bigg] ~~.
\end{align}
Just as before, we now find the values of $(a_{M,4},a_{M,2})$ and $(a_{S,3},a_{M,2})$ which minimize the degree of EoS variability. Figure~\ref{fig:fracdiff2324n0n1} shows contours of fixed relative fractional difference in the $(a_{M,4},a_{M,2})$ (left panel) and $(a_{S,3},a_{M,2})$ (right panel) planes. Observe again that there are one-parameter families for which the fractional difference is exactly zero. This time, however, the original normalizations of~\cite{Yagi:2014bxa} are not quite on these one parameter families. This is important because it indicates that there are better choices of normalization that minimize the EoS variability further. Also as before, observe that the EoS variability increases as one chooses normalization values away from these one-parameter families. Moreover, there is a line in the $(a_{M,4},a_{M,2})$ and $(a_{S,3},a_{M,2})$ planes for which the fractional differences are not well-defined.  

\begin{figure*}[htb]
\begin{center}
\includegraphics[width=8.0cm,clip=true]{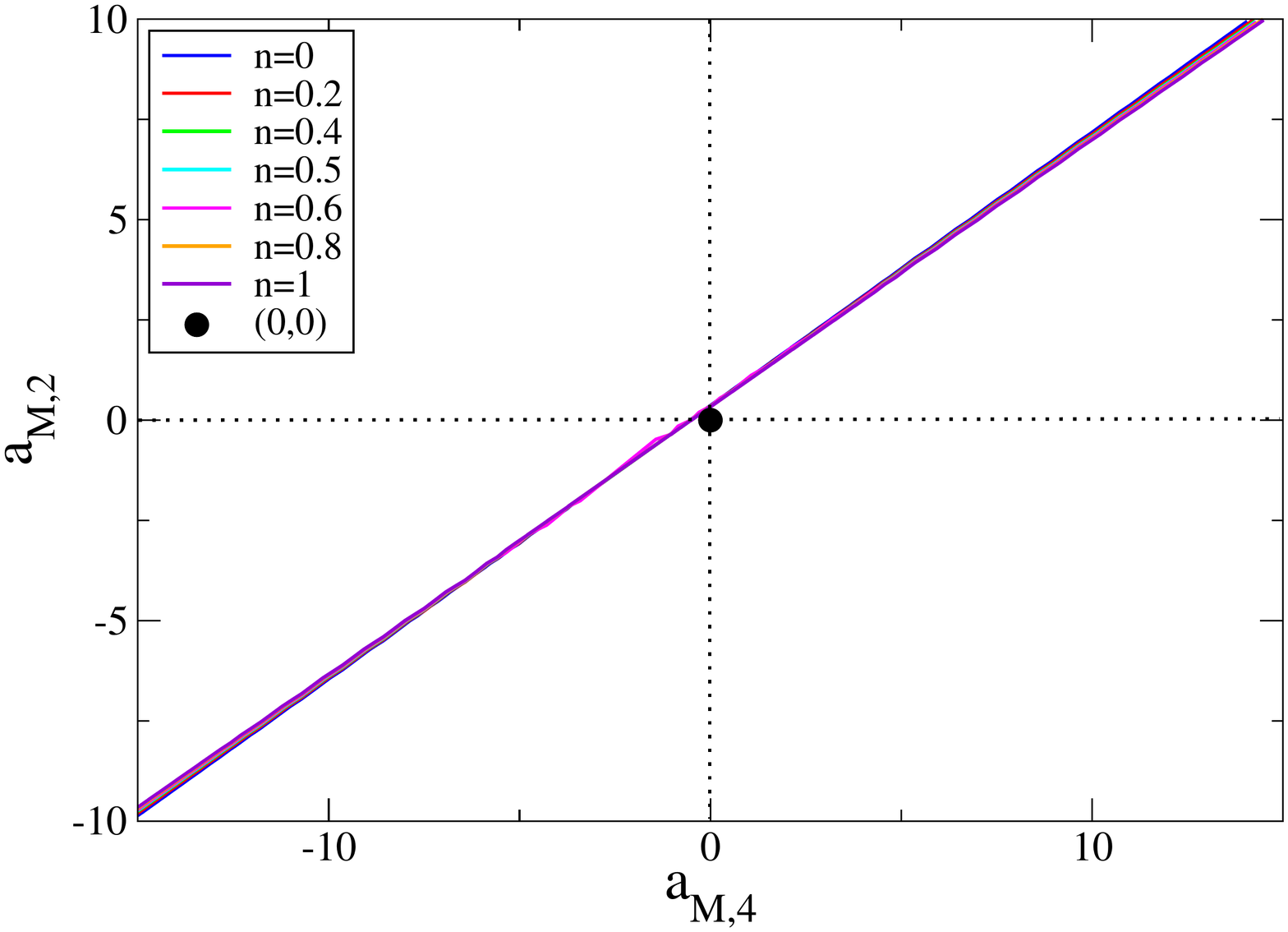} 
\hspace{0.5cm}
\includegraphics[width=8.0cm,clip=true]{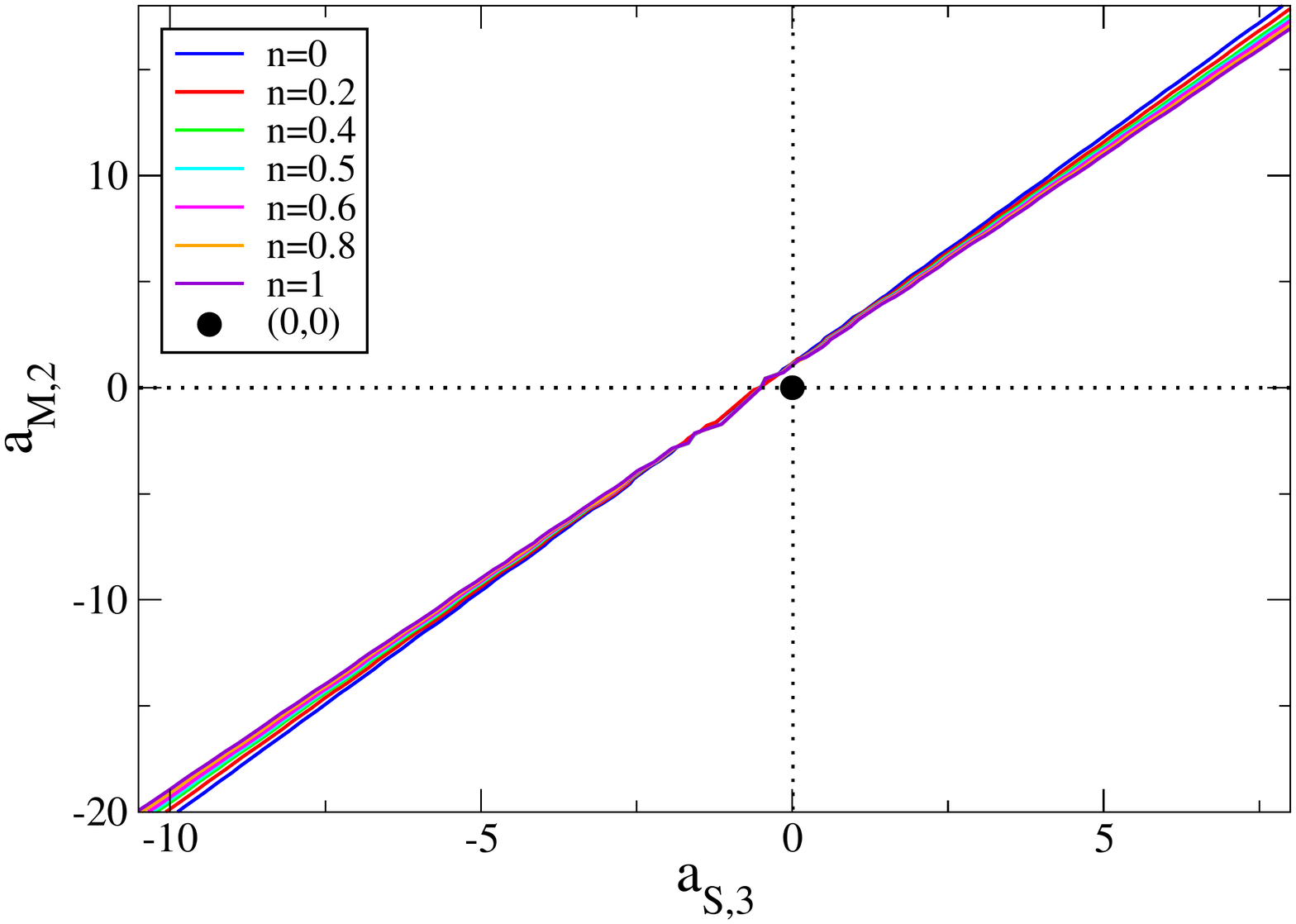}  
\caption{\label{fig:vlinem4s3} (Color Online) Same as Fig.~\ref{fig:LEfrac} but using the $M_4$-$M_2$ in the $(a_{M,2},a_{M,4})$ plane (left) and $S_{3}$-$M_{2}$ in the $(a_{S,3},a_{M,2})$ plane (right) relations. Observe again that all the curves cluster around each other and that the original normalizations of~\cite{Yagi:2014bxa} do not quite coincide with the normalizations that lead to the least EoS variability.}
\end{center}
\end{figure*}

As in the I-Q case, one may wonder whether these one-parameter families change if one computes the relative fractional differences in the $\bar{M}_4$-$\bar{M}_2$ and $\bar{S}_3$-$\bar{M}_2$ between other EoSs. Let us then repeat the study carried out in the I-Q case with polytropic EoSs. Figure~\ref{fig:vlinem4s3} shows the one-parameter families for which the relative fractional difference in the $\bar{M}_4$-$\bar{M}_2$ and $\bar{S}_3$-$\bar{M}_2$ relations is exactly zero between a polytropic EoS with index $n$ and a reference polytropic EoS with $n = 0.643$. Observe that all the one-parameter families cluster around single curves. Moreover, the original normalization chosen in~\cite{Yagi:2014bxa} does not lie on any of these families. This suggests that there may be an optimal one-parameter family of normalizations that is different from the original choice of~\cite{Yagi:2014bxa} which improves the EoS universality of the $\bar{M}_4$-$\bar{M}_2$ and $\bar{S}_3$-$\bar{M}_2$ relations.  
%

\section{EoS Universality \\ in Full General Relativity}
\label{sec:GR}

We have seen that it is possible to construct I-Q, $M_4$-$M_2$ and $S_3$-$M_2$ relations that are properly normalized such that they are essentially EoS insensitive in the Newtonian limit and for simple polytropic EoSs; but does such a choice of normalization also lead to EoS insensitivity in the relativistic regime and for realistic EoSs? This section is devoted to answering this question. 

Let us start by reviewing how the I-Q, $M_4$-$M_2$ and $S_3$-$M_2$ relations are computed in full GR, following~\cite{I-Love-Q-PRD,Yagi:2014bxa,Kojima:1999jk}. We construct solutions to the Einstein equations that represent isolated, unmagnetized and slowly rotating NSs, perturbatively  in the ratio of the spin angular momentum to the stellar mass squared with the Hartle-Thorne framework~\cite{hartle1967,hartlethorne}. In order to compute the hexadecapole moment $M_{4}$, we must retain terms up to fourth order in the small-rotation parameter.  The matter sector is modeled as a perfect fluid with realistic EoSs, such as APR \cite{APR}, SLY \cite{SLy,shibata-fitting}, LS220 \cite{LS,ott-EOS}, Shen \cite{Shen1,Shen2}, WFF1 \cite{Wiringa:1988tp} and ALF2 \cite{Alford:2004pf}. All these EoSs support NSs with masses greater than $2M_{\odot}$, which is needed given the two recently discovered massive pulsars~\cite{2.01NS,1.97NS}. We also include a few simulations with polytropic EoSs for comparison with the previous section. 

The equations of structure are then solved order by order in the slow-rotation expansion  (see e.g.~\cite{I-Love-Q-PRD,Kojima:1999jk,Yagi:2014bxa}). At any given order in rotation, one must solve the structure equations numerically in the interior of the star and then match them at the stellar surface to an exterior solution. The numerical calculations are done with an adaptive $4^{th}$ order Runge-Kutta integrator \cite{galassi2009gnu}. The boundary conditions at the stellar center are obtained through a local analysis of the structure equations at the stellar core, while the boundary conditions at spatial infinity are fixed via asymptotic flatness. The matching of the solutions at $N$th order in rotation yields the $N$th multipole moment of the NSs.

\begin{figure}[htb]
\begin{center}
\includegraphics[width=8.0cm,clip=true]{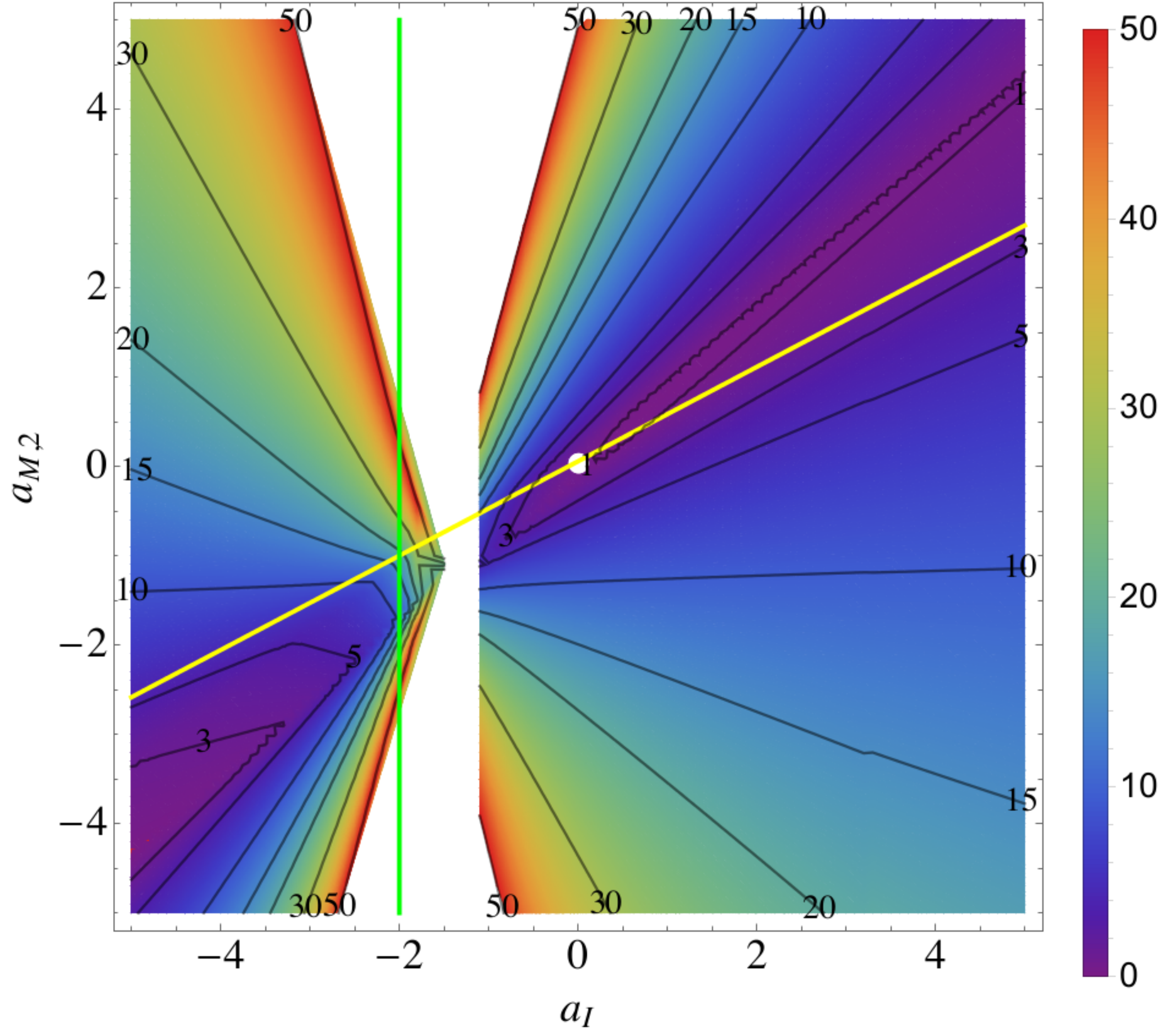}
\caption{\label{fig:CP_IQ_full} (Color Online) 
Same as left panel of Fig.~\ref{fig:CP_IQ_half} but zoomed out.  Observe that the original normalization of~\cite{I-Love-Q-PRD,I-Love-Q-Science} is in the region of $(a_{M,2},a_{I})$ that lead to the least EoS-variability. Observe also that the one-parameter family of normalizations that lead to the least EoS variability in a Newtonian treatment (yellow line) does not coincide with the regions that lead to the least EoS variability in full GR (purple regions). Finally, observe also that the region that leads to the most EoS variability (white region) is shifted from what one would expect from a Newtonian analysis (green line). 
}
\end{center}
\end{figure}

We now investigate the \emph{no-hair} relations in the relativistic regime and with realistic EoSs, using the numerical framework described above. Let us first focus on the I-Q relations and let us normalize the moment of inertia and the quadrupole moment as in Eqs.~\eqref{eq:I-Norm} and~\eqref{Meqn}. These relations then become a function of $(a_{M,2},a_{I})$ and we wish to determine the set that minimizes the degree of EoS variability. To do so, we will compute the relative fractional difference in the I-Q relations with different realistic EoSs, with respect to the relations computed with a \emph{reference} EoS, which in this case we take to be LS220. Figure~\ref{fig:CP_IQ_full} shows the contours of \emph{maximum} relative fractional difference in the I-Q relations in the $(a_{M,2},a_{I})$ plane, i.e.~for a discrete set of values in the $(a_{M,2},a_{I})$ plane, we compute the relative fractional differences in the I-Q relations calculated with an LS220 EoS and all other realistic EoSs, and then, we create contours of the maximum values of those relative fractional differences.  Observe that there still exists a set of normalizations for which the maximum EoS variation is $\sim$ 1\%, as shown by the violet regions of the figure. Observe also that the original normalization used in~\cite{I-Love-Q-PRD}, denoted by a white dot in Fig.~\ref{fig:CP_IQ_full}, is close to the best choice of normalization. The one-parameter family of normalization that gives the least EoS variability in the Newtonian case (yellow line) disagrees with that which minimizes the EoS variability in the relativistic case. Yet still, in the relativistic case, it seems like there is a one-parameter family that minimizes the EoS variability, except that now the relativistic corrections modify the slope of the Newtonian relation. Another important observation is that there exist choices of normalization (white region) for which the maximum EoS variation is not well behaved numerically; this agrees with the divergent region (green line) in the Newtonian case. A zoomed-in version of the first quadrant of this figure is shown in Fig.~\ref{fig:CP_IQ_half} (left panel). In conclusion, we found a family of normalizations that lead to strong EoS insensitivity in the I-Q relations, which includes the original normalization of~\cite{I-Love-Q-Science,I-Love-Q-PRD}, in the relativistic regime with realistic EoSs.  

\begin{figure}[htb]
\begin{center}
\includegraphics[width=8.0cm,clip=true]{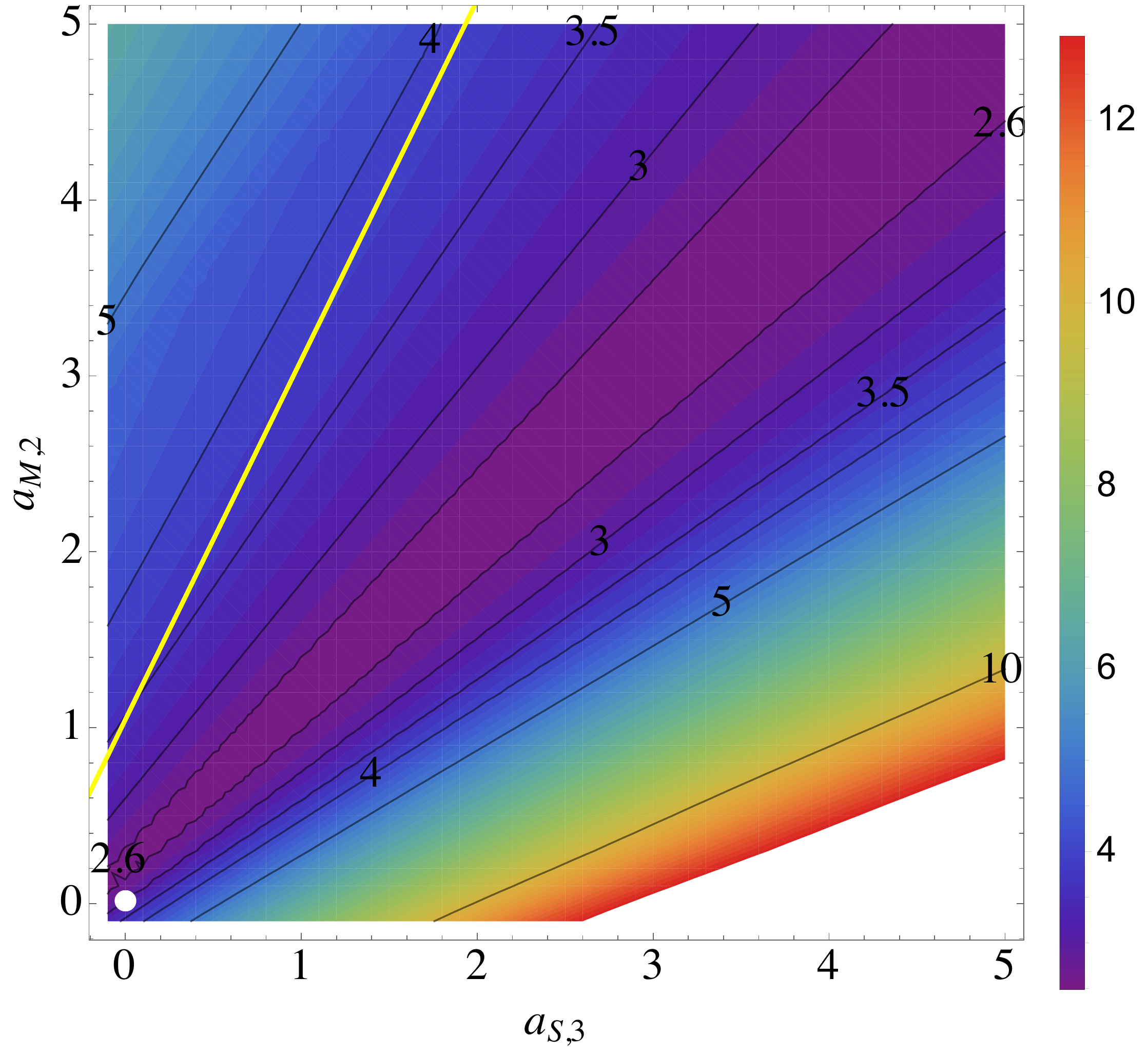}
\caption{\label{fig:CP_S3M2_half} (Color Online) 
Same as right panel of Fig.~\ref{fig:CP_IQ_half} but for the $S_3$-$M_2$ relation in the $(a_{S,3},a_{M,2})$ plane. Observe again that the one-parameter family of normalizations that minimizes the EoS variability in the Newtonian regime (yellow line) does not coincide with the set of normalizations that does so in full GR (purple regions). Importantly, observe that the original normalization chosen in~\cite{Yagi:2014bxa} is quite far from the purple region. This implies that there are better choices of normalization that would make the $S_{3}$-$M_{2}$ relations more EoS universal.}
\end{center}
\end{figure}

Let us now investigate higher multipole order, no-hair relations and attempt to determine the normalizations that lead to the strongest EoS universality. In particular, let us focus on the $\bar{M}_4$-$\bar{M}_2$ and $\bar{S}_3$-$\bar{M}_2$ relations in the relativistic regime and with realistic EoSs. Figures~\ref{fig:CP_IQ_half} (right panel) and~\ref{fig:CP_S3M2_half} show the contours of maximum relative fractional difference in these relations due to EoS variation in the plane of the normalization parameters. Observe that in both cases there is a region in the normalization plane that lead to the least EoS variability and that resembles a one-parameter family.  
Observe also that this one-parameter family in the relativistic case is different from the one found in the Newtonian case (yellow lines), just as in the I-Q relations. This time, however, the Newtonian one-parameter family seems to be close to its relativistic version. 
Importantly, observe that the original normalizations in~\cite{Stein:2013ofa,Yagi:2014bxa} are not that close to the relativistic one-parameter family that minimizes the EoS variability. For example, the best choice of normalization for the $M_4-M_2$ relations leads to a maximum EoS variation of $\sim$ 6\%~\cite{Yagi:2014bxa}, while the maximum EoS variation using the original normalization was $\sim$ 9\%. Similarly, the best choices of normalization for the $S_3-M_2$ relations lead to a maximum EoS variation of $\sim$ 2\%, while the original parameterization lead to a variation of 3.5\%~\cite{Yagi:2014bxa}. 
In conclusion, we have found the set of normalization parameters that minimizes the EoS variability, and this set is different from the original normalization used in~\cite{Yagi:2014bxa}, enhancing universality by a factor of 2-4. 

\section{Discussion}
\label{sec:conclusion}

The recently-discovered, approximately EoS insensitive, I-Love-Q relations are important for future observations. In particular, the I-Q relations are helpful in determining the mass and radius of NSs from the X-ray pulse profile emitted by millisecond pulsars. The NICER and LOFT collaborations hope to measure the mass and radius to 5\% accuracy, and thus, one requires I-Q relations that are EoS-insensitive to at least this level. Moreover, the expected pulse profile may also be affected by higher-order-in-spin corrections, for example induced by the octopole and hexadecapole moments of the NS, provided the latter is spinning sufficiently fast~\cite{Yagi:2014bxa}. Thus, one would wish to also find $M_4$-$M_2$ and $S_3$-$M_2$ relations that are the least EoS sensitive as possible.    

In this paper, we have studied how to construct I-Q and higher multipole order, no-hair relations for NSs that maximize the degree of EoS insensitivity. To do so, we investigated whether one can normalize these quantities in a way that strengthens the EoS insensitivity relative to the original normalizations used in~\cite{I-Love-Q-PRD}. We have found the optimal one-parameter family of normalizations that leads to the strongest EoS-insensitivity both with a Newtonian and a fully relativistic analysis. We have observed that this one parameter family is strongly affected by relativistic corrections, and the best normalization in the Newtonian limit does not necessarily lead to the least EoS variation in the relativistic regime. Yet,regarding the I-Q relations, there is little gain in EoS insensitivity when using these normalizations relative to the original ones of~\cite{I-Love-Q-PRD}, because the latter happened to coincidentally lie very close to the optimal family.  
On the other hand, relativistic corrections to the $M_4$-$M_2$ and $S_3$-$M_2$ relations greatly affect this one-parameter family; in fact, there are choices that can improve the degree of EoS-insensitivity relative to the original normalizations of \cite{Yagi:2014bxa}. 

One could extend the present study to other modified theories of gravity, such as dynamical Chern-Simons gravity (DCS)~\cite{CSreview,Yagi:2013mbt}. Modified theories typically introduce dimensional coupling constants that characterize the degree to which they differ from GR. One could thus study what normalization minimizes the degree of EoS variability in such modified theories.

Quite recently the I-Q relations were studied for highly magnetized NSs \cite{I-Love-Q-B}. The authors showed that for strong magnetic fields ($B>10^{12}$G) with a twisted-torus configuration the relations lose universality. The NSs considered in \cite{I-Love-Q-B} are characterized by dimensional magnetic fields but we know that the universality of the I-Q relation depends heavily on how we normalize the NS observables. Future work could concentrate on investigating whether the universality holds for highly magnetized NSs if a different choice of normalization for the magnetic field strengths is made.

\section{Acknowledgements}
NY acknowledges support from NSF CAREER Award PHY-1250636. BM is supported by the Fulbright-Nehru Postdoctoral Research Fellowship. Some calculations used the computer algebra-systems \textsc{MAPLE}, in combination with the \textsc{GRTENSORII} package~\cite{grtensor}.

\bibliography{master}
\end{document}